\documentclass{aastex63}
\usepackage{subfigure}
\usepackage{amsmath}
\usepackage{comment}
\usepackage{xcolor}
\usepackage{colortbl}
\usepackage{multirow}
\usepackage{subfiles} 

\chardef\us=`\_

\received{January 6, 2021}
\revised{April 11, 2021}
\accepted{\today}
\submitjournal{ApJ}

\shorttitle{Physics-driven machine learning for CMEs}
\shortauthors{Guastavino et al.}
\graphicspath{{./}{figures/}}

\begin{document}

\title{Physics-driven machine learning for the prediction of coronal mass ejections' travel times}

\correspondingauthor{Michele Piana}
\email{piana@dima.unige.it}

\author{Sabrina Guastavino}
\affiliation{MIDA, Dipartimento di Matematica, Università di Genova, via Dodecaneso 35 16146 Genova, Italy}
\author{Valentina Candiani}
\affiliation{MIDA, Dipartimento di Matematica, Università di Genova, via Dodecaneso 35 16146 Genova, Italy}
\author{Alessandro Bemporad}
\affiliation{Istituto Nazionale di Astrofisica (INAF), Osservatorio Astrofisico di Torino}\author{Francesco Marchetti}
\affiliation{Dipartimento di Matematica \lq\lq Tullio Levi-Civita", Università di Padova, Padova, Italy}
\author{Federico Benvenuto}
\affiliation{MIDA, Dipartimento di Matematica, Università di Genova, via Dodecaneso 35 16146 Genova, Italy}
\author{Anna Maria Massone}
\affiliation{MIDA, Dipartimento di Matematica, Università di Genova, via Dodecaneso 35 16146 Genova, Italy}
\author{Roberto Susino}
\affiliation{Istituto Nazionale di Astrofisica (INAF), Osservatorio Astrofisico di Torino}
\author{Daniele Telloni}
\affiliation{Istituto Nazionale di Astrofisica (INAF), Osservatorio Astrofisico di Torino}
\author{Silvano Fineschi}
\affiliation{Istituto Nazionale di Astrofisica (INAF), Osservatorio Astrofisico di Torino}
\author{Michele Piana}
\affiliation{MIDA, Dipartimento di Matematica, Università di Genova, via Dodecaneso 35 16146 Genova, Italy}
\affiliation{Istituto Nazionale di Astrofisica (INAF), Osservatorio Astrofisico di Torino}



\begin{abstract}

Coronal Mass Ejections  (CMEs) correspond to dramatic expulsions of plasma and magnetic field from the solar corona into the heliosphere. CMEs are scientifically relevant because they are involved in the physical mechanisms characterizing the active Sun. However, more recently CMEs have attracted attention for their impact on space weather, as they are correlated to geomagnetic storms and may induce the generation of Solar Energetic Particles streams. In this space weather framework, the present paper introduces a physics-driven artificial intelligence (AI) approach to the prediction of CMEs travel time, in which the deterministic drag-based model is exploited to improve the training phase of a cascade of two neural networks fed with both remote sensing and in-situ data. This study shows that the use of physical information in the AI architecture significantly improves both the accuracy and the robustness of the travel time prediction.

\end{abstract}


\section{Introduction}

Coronal Mass Ejections (CMEs) \citep{2011ASSL..376.....H} consist in large eruptions of plasma and magnetic field that are typically triggered by solar flares \citep{piana2022hard} and that propagate from the solar corona into the heliosphere. From a phenomenological perspective, the rate of occurrence of CMEs is related to the solar cycle and typically ranges from one event per more than one week at the minimum to several CMEs per day at the maximum \citep{Zhao2014}. From an experimental perspective, the observations of CMEs are typically performed by means of remote-sensing instruments that can measure their most significant kinematic parameters, like 
the initial propagation speed, 
the CME mass and initial cross-section. 
Examples of telescopes appropriate for measuring remote sensing parameters are coronagraphs on board space clusters like the Large Angle and Spectrometric Coronagraph (LASCO) \citep{1995SoPh..162..357B} on board the Solar and Heliophysics Observatory (SOHO) \citep{1995SoPh..162....1D}, the Sun Earth Connection Coronal and Heliospheric Investigation (SECCHI) \citep{2008SSRv..136...67H} on board STEREO-A/STEREO-B \citep{2008SSRv..136....5K}, and the recent Metis \citep{fineschi2012metis} on board Solar Orbiter \citep{muller2020solar}. Further, CMEs travel from the Sun to the Earth while embedded within the solar wind \citep{2012esw..book.....L}, which implies that some solar wind parameters play a significant role for the understanding of the CMEs dynamics. In particular, the solar wind average density and speed can be inferred from measurements provided by in-situ instruments like the WIND Spacecraft \citep{2021RvGeo..5900714W}, the Advanced Composition Explorer (ACE) \citep{1998SSRv...86....1S}, the Charge, Element, and Isotope Analysis System (CELIAS) \citep{hovestadt1995celias} on board SOHO, and the Solar Wind Analyzer (SWA) \citep{owen2020solar} on board Solar Orbiter. 
 
Besides their relevance for the comprehension of the physical mechanisms involved in the active Sun, CMEs have recently attracted notable attention also in the space weather context \citep{gopalswamy2009coronal,howard2014space}, and this is due to several reasons. First, CMEs are often correlated with the occurrence of geomagnetic storms when interacting with Earth's magnetosphere; second, fast CMEs may induce interplanetary shocks that may contribute to the generation of intense Solar Energetic Particles (SEPs) streams; finally, and most importantly, when directed toward the earth, CMEs may impact the correct functioning of both space- and ground-based communication, navigation, and energy production systems.

The study of the space weather impacts of CMEs implies the need of formulating, implementing, and validating forecasting approaches that must have the potentiality to easily become operational services for space weather monitoring and nowcasting. Specifically, the typical space weather end-user is primarily interested in rather practical issues like whether or not a CME that has been detected in the corona by a remote-sensing telescope will hit the Earth; and, if so, at which time (Time of Arrival, ToA) and speed (Speed of Arrival, SoA) the impact will occur. Focusing on the prediction of the ToA, these forecasting problems have been addressed in several fairly 
recent studies utilizing computational approaches that can be clustered into three families \citep[see][and references therein]{Zhao2014}. Empirical models \citep{gopalswamy2000interplanetary} adopt simple equations to fit the relationship between the CME travel time (TT) and the corresponding observed parameters at the Sun; physics-based models \citep[see][and references therein]{pomoell2018euhforia} introduce physics to describe the CME propagation and, in the magnetohydrodynamic (MHD) versions, they can even account for the state of the background heliosphere; finally, machine/deep learning (ML/DL) approaches rely on artificial intelligence (AI) purely data-driven algorithms to estimate the ToA given large sets of observations of the CME parameters at the Sun \citep{Sudar2016,Liu2018,Shi2021} or CME images \citep{Wang2019,Fu2021,Alobaid2022}.

A reliable assessment of the prediction effectiveness of such methods is difficult for at least two reasons \citep{Camporeale2019,Vourlidas2019}. First, these studies are performed using data acquired by means of different instruments and the lack of data standardization can impact the reliability of the prediction. Second, both the computational conditions under which the experiments are implemented and the ways the prediction effectiveness has been evaluated are often significantly heterogeneous. An effort to compare results obtained by different prediction models is being made by the NASA's Community Coordinated Modeling Center (CCMC) (\url{https://kauai.ccmc.gsfc.nasa.gov/CMEscoreboard/}) and a first extensive discussion of the obtained results is contained in \cite{Riley2018}. Both the tables obtained in that study and the ones contained in more recent applications of AI-based techniques \citep[see][and references therein]{camporeale2019challenge} provide results that are significantly poor as far as both the ToA prediction accuracy and its robustness are concerned (on average, the prediction errors are almost systematically larger than $10$ hours, with standard deviations that may exceed 20 hours).

In the present paper we aim at improving these prediction estimates by following an approach that combines the computational effectiveness of AI with physical information contained in deterministic models. Conceptually, such kind of combination can be done in two ways: either data-driven AI can be used to constrain the parameters contained in the MHD equations, or physics-based models can be exploited to better realize the training phase in supervised ML/DL processes. Following this latter path, here we used the well-established drag-based model \citep{Cargill2004,Vrsnak2010,Vrsnak2013} to design the loss functions of an architecture made of two neural networks, each one characterized by six hidden layers. Among the well-established approaches based on kinematic models, which rely on simplified assumptions on the solar wind propagation mechanisms and on the interaction between solar wind and CMEs, the drag-based model is mostly used thanks to its notable computational effectiveness and the limited number of input parameters. Specifically, the only input parameter of the model that is not provided by experimental measurements and that, therefore, must be a priori estimated is the free drag parameter. The first neural network of our AI architecture is applied to estimate such parameter and is characterized by a fully model-driven loss function. Then, the second network, which has a loss function that is a weighted combination of a data-driven and a model-driven component, is used to predict the CME TT. The results we obtained by means of this physics-supported AI approach showed that 1) the accuracy with which the free drag parameter is estimated has a notable impact on the accuracy of the TT forecast; and, 2) an AI architecture that incorporates the deterministic physical model in the training process performs better than a purely data-driven algorithm in term of both the accuracy and the robustness of the prediction.

The plan of the paper is as follows. Section 2 provides a quick overview of the drag-based model. Section 3 illustrates the neural network architecture designed for the analysis. Section 4 shows the results of our approach when applied to a limited set of LASCO observations. Our conclusions are offered in Section 5.

\section{The drag-based model}

According to drag-based models \citep{vrvsnak2013propagation,vzic2015heliospheric,dumbovic2021drag}, the kinematics of CMEs is mainly determined by their interaction with the solar wind in which they are immersed. Specifically, the origin of the name of such approaches is in the fact that the drag acceleration (or deceleration) must follow a fluid dynamic analogy, i.e., it must have a quadratic dependence on the relative speed between the CME and the background solar wind. As a consequence, the standard drag-based model equation reads as
\begin{equation}\label{eq:model}
    \ddot{r}(t) = -\gamma |\dot{r}(t)-w|(\dot{r}(t) - w) \ ,
\end{equation}
where $r(t)$, $\dot{r}(t)$, and $\ddot{r}(t)$ are the position, speed, and acceleration of the CME as a function of time $t$, respectively; $w(r,t)$ is the solar wind speed and $\gamma$ is the drag parameter, which measures the interaction effectiveness between the CME and the solar wind and it can be expressed as
\begin{equation}\label{eq:drag-parameter}
    \gamma =  C \frac{A\rho}{m},
\end{equation}
where $A$ and $m$ are the CME impact area and mass, respectively; $\rho(r,t)$ is the solar wind density; $C$ is the dimensionless dimensionless drag coefficient. Equation (\ref{eq:model}) is completed to a Cauchy problem by including the two initial conditions $r(0)=r_0$ and $\dot{r}(0)=v_0$, where $r_0$ is the height of the eruption ballistic propagation, and $v_0$ is the initial CME speed. Both $r_0$ and $v_0$ should be considered known and provided as experimental measurements. Physical limitations of the drag-based model have been fully described in \citet{Vrsnak2013}. In particular, the model considers a simplified structure for the background solar wind, i.e., it is assumed that all parts of the CME are embedded in an isotropic flow, where the flow speed does not change with distance. 

A typical application of the drag-based model is the estimate of the CME ToA given estimates and measurements of the model parameters. In fact, assuming in the following that the solar wind speed and the drag parameter are constant and homogeneous, equation (\ref{eq:model}) leads to 
\begin{equation}\label{eq:ICME_speed}
    \dot{r}(t) = \frac{v_0-w}{1 + \gamma \ \mathrm{sign}(v_0-w)  (v_0-w)t} + w \ ,
\end{equation}
and
\begin{equation}\label{solution-drag}
     r(t)=  \frac{1}{\gamma} \mathrm{sign}(v_0-w) \log\left(1 + \gamma  \mathrm{sign}(v_0-w) (v_0-w) t \right) + wt + r_0.
\end{equation}
Equation (\ref{solution-drag}) can be used to estimate the TT as the solution of $r(t)=1$ AU. Once the TT is estimated, it can be included in equation (\ref{eq:ICME_speed}) to obtain an estimate of the SoA. If we substitute $\gamma$ with equation \eqref{eq:drag-parameter} 
this approach is reliable just if accurate estimates of the parameters $A$, $m$, $C$, $\rho$, $w$, $r_0$, and $v_0$ are at disposal. Specifically, measurements of $A$, $m$, $r_0$, and $v_0$ can be provided by coronagraphic instruments like LASCO or, more recently, Metis on board Solar Orbiter. In situ instruments like WIND, ACE, and CELIAS can provide estimates of $\rho$ and $w$ at the CME onset and these same values can be used in the equations as approximations of average values of the two parameters. The determination of $C$ is particularly critical. A possible procedure is given in \citet{Napoletano2018} and another approach is described below, together with a discussion of the impact of the accuracy of this estimate on the determination of the TT.

\section{Neural networks architecture}
We propose an AI-based, physics-supported approach to the estimate of the CME TT that exploits the use of two neural networks in cascade (see Figure \ref{fig:NN-cascade}). The two algorithms have the same design, whose parameters are contained in Table \ref{tab:NN-cascade-parameters}. The first network ($N1$) can take as input measurements of the initial CME speed ($v_0$), the CME mass ($m$) and impact area ($A$), together with estimates of the solar wind density ($\rho$) and speed ($\omega$). The output of this first network is an estimate of the drag parameter $C$. Therefore, the second neural network ($N2$) can take as input the same parameters as the first network plus this estimate of $C$ and forecasts the corresponding TT (we point out that the CME ToA is easily determined from the corresponding TT, since the time of occurrence of the CME at onset is known from observations).

\begin{table}[]
\centering
\caption{Network settings for both neural network architectures. The only difference lies in the number of values in the input layer, $5$ for the first network N1 and $5$ or $6$ for the second network $N2$, depending on the considered configuration (cf. Table \ref{tab:configurations}). The activation function is the Rectified Linear Unit (ReLU) $f(x)=\max(0,x)$ for the input and the hidden layers, and the SoftPlus function $f(x)=\log(1+\exp(x))$ for the output.}
\setlength{\tabcolsep}{8pt}
\begin{tabular}{l|| r| r| r| r| r| r| r| r }
\hline
Layer               & Input & Hidden 1    & Hidden 2    & Hidden 3    & Hidden 4    & Hidden 5    & Hidden 6    & Output   \\ 
Nodes               & 5/6   & 200  & 100  & 50   & 30   & 25   & 10   & 1        \\ 
Activation function & ReLU  & ReLU & ReLU & ReLU & ReLU & ReLU & ReLU & SoftPlus \\ \hline
\end{tabular}
\label{tab:NN-cascade-parameters}
\end{table}

\subsection{Loss functions}

The training phase for the two networks relies on choices of the loss functions that, programmatically, want to incorporate physical information encoded into the drag-based model. To accomplish this objective, the loss function must be differentiable. Therefore the sign function at the denominator of (\ref{eq:ICME_speed}) is approximated as
\begin{equation}\label{eq:approx_sign}
    \mathrm{sign}(v_0-w) \approx \frac{(v_0-w)}{\sqrt{(v_0-w)^2 + \delta}},
\end{equation}
where $\delta$ is a small positive value.
By incorporating approximation (\ref{eq:approx_sign}) into (\ref{eq:ICME_speed}), by analytically integrating the resulting form for $\dot{r}(t)$ and by substituting $\gamma$ with equation \eqref{eq:drag-parameter}, we obtained
\begin{equation}\label{solution-drag-approx}
     r(t,C)=  \frac{1}{\frac{A}{m}C\rho\sigma}\log\left(1+ \frac{A}{m}C\rho \sigma  (v_0-w) t \right) + wt + r_0,
\end{equation}
where
\begin{equation}\label{sigma}
\sigma = \frac{(v_0-w)}{\sqrt{(v_0-w)^2 + \delta}} \ .
\end{equation}

The first network $N1$ estimates the drag parameter $C$ by utilizing the fully model-inspired quadratic loss function
\begin{equation}\label{eq:loss-C}
L_c(t,N1(\mathbf{x})) = \left( r(t,N1(\mathbf{x})) - 1 \right)^2 = \left(\frac{m \sigma}{A \rho N1(\mathbf{x})} \log\left(1 + \sigma \frac{A \rho N1(\mathbf{x})}{m} (v - w)t + w t \right) + r_0 -1\right)^2 ,
\end{equation}
where $\mathbf{x}=(v_0,m,A,\rho,\omega)$ is the input vector and the CME position is measured in astronomical units. We point out that, in this first case, observational values are used for $t$, but the estimate of the drag-parameter provided by the network $N1$ is not intended as explicitly depending on this time values ($N1$ depends just on the five input parameters). In this way, once the network is trained, it is able to forecast $C$ without the need of any knowledge of $t$, which is the actual unknown quantity to predict at the end of the whole neural network cascade.
\\
The second network $N2$ predicts the TT and, in this case, the loss function is a weighted sum of a fully data-driven quadratic function and of a fully model-driven component, i.e.,
\begin{equation}\begin{split}\label{eq:loss-ToA}
L_{t}(t,N2(\bar{\mathbf{x}})) &=
\lambda (t - N2(\bar{\mathbf{x}}) )^2 + (1-\lambda) (r(N2(\Bar{\mathbf{x}}),N1(\textbf{x}))-1)^2 \\
& = \lambda (t - N2(\bar{\mathbf{x}}) )^2 + (1-\lambda) \left(\frac{m \sigma}{A \rho N1(\textbf{x})} \log\left(1 + \sigma \frac{A \rho N1(\textbf{x})}{m}\right) (v - w)N2(\bar{\mathbf{x}}))+ v N2(\Bar{\mathbf{x}})) + r_0 -1\right)^2 ,
\end{split}
\end{equation}
where $N1(\mathbf{x})$ is the value predicted by the first neural network, $N2(\Bar{\mathbf{x}})$ is the predicted TT and $\bar{\mathbf{x}}$ is the input vector for $N2$, which may or may not contain the estimate of $C$ provided by $N1$. 

\subsection{Cascade configurations}
\begin{figure}
\begin{center}
\includegraphics[width=16.cm]{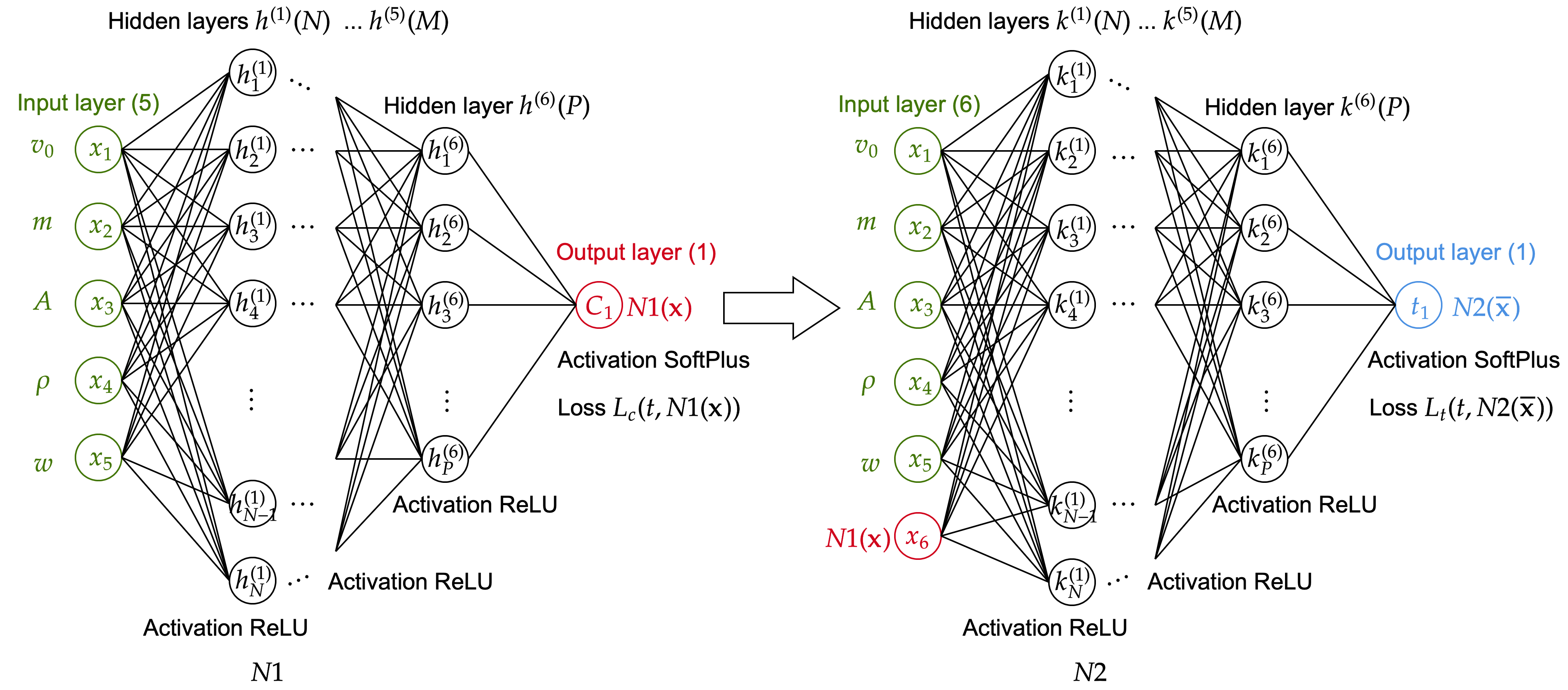}
\caption{Neural network architectures in cascade: the first network $N1$ can be used to estimate the drag parameter $C$ from CME data. The second network $N2$ is then employed to estimate the ToA of the considered CME at 1 AU. Here $N=200$, $M=25$ and $P=10$ are the number of neurons in the first, fifth and sixth hidden layers, respectively.}
\label{fig:NN-cascade}
\end{center}
\end{figure}

The design in Figure \ref{fig:NN-cascade} is flexible enough to allow the TT prediction according to the six possible configurations described in Table \ref{tab:configurations}. The differences characterizing these configurations depend on the role played by the drag-based model in the training phase of $N2$, and by the fact that the drag parameter $C$ either may or may not be used as input feature. Specifically, in Configuration 1 (C1), Configuration 2 (C2), and Configuration 3 (C3), the input of the second network is $\Bar{\mathbf{x}} = \mathbf{x}$, i.e., the drag-parameter is not used as input feature. In C1, $N1$ is switched off, while the loss function in $N2$ is fully data-driven (i.e., $\lambda=1$). In configurations C2 and C3 the first network is switched on and estimates of $C$ are used in the loss function of $N2$: indeed, in these two latter configurations the loss function is represented by the weighted sum of a fully data-driven and a fully physics-driven component (i.e., $\lambda=0.5$), and is fully physics-driven (i.e., $\lambda=0$), respectively. However, in C2 and C3 the second network does not use $C$ as an input feature. In the remaining three configurations, the first network is always switched on and the drag parameter $C$ is always utilized as a feature of the second network, with values provided as predictions by the first network. In particular, in Configuration 4 (C4), Configuration 5 (C5), and Configuration 6 (C6), the loss function of $N2$ is fully data-driven (i.e., $\lambda=1$ in $N2$), has both the data- and physics-driven components (i.e., $\lambda=0.5$), and is fully physics-driven, respectively.

\begin{table}[]
\centering
\caption{In the first three configurations C1, C2, C3 the drag parameter $C$ is not considered as an input of the second neural network $N2$, and it is estimated by the first network $N1$ only when needed in the loss function $L_t$. The choice of the parameter $\lambda = 1$ corresponds to the fully data-driven case, whereas $\lambda = 0.5$ and $\lambda = 0$ represent the mixed and the fully model-driven case, respectively.}
\begin{tabular}{c|ccc|c}
\multirow{2}{*}{Configuration} & \multicolumn{3}{c|}{Training phase} & \multicolumn{1}{c}{Testing phase} \\ \cline{2-5} 
 & $N1$      & $N2$      & $\lambda$     & drag parameter as input of $N2$      \\ \hline
C1 & off & on & 1 & off  \\
C2 & on & on & 0.5 & off  \\
C3 & on & on & 0 & off \\  \hline
C4 & on & on & 1 & on \\
C5 & on & on & 0.5 & on  \\
C6 & on & on & 0 & on  \\ \hline 
\end{tabular}
\label{tab:configurations}
\end{table}

\subsection{Optimization}
The two networks have been trained over 10000 epochs using the Adam optimizer and learning rate equal to $10^{-3}$ \citep{Kingma14}. The validation of the networks has been performed by using an early stopping rule based on the monitoring of the loss function in the validation phase \citep{caruana2000overfitting}. Specifically, we stop the iterations when the loss function during validation does not improve after 2000 epochs. The two networks are characterized by six hidden layers, and we initialized the network weights with a random uniform distribution in the range $(0, 0.01)$. Details of the two networks are provided in Table \ref{tab:configurations}.

\newpage
\section{Applications}

\subsection{Dataset}
We have performed the validation of this approach to TT forecasting by means of a set of 123 events observed by LASCO \citep{DelMoro2022} and CELIAS. The events occurred in the time range between 1997 and 2018 (\citep{Richardson2010}, hereafter dubbed R \& C). Specifically, LASCO provided all the CME parameters at the onset, while CELIAS provided the solar wind density and speed at the onset (see Table \ref{tab:dataset}). These observed values have been used as $\rho$ and $w$ in the loss functions, i.e., we assumed the homogeneity and the stationarity of these solar wind parameters. A pre-processing step was necessary to filter out the events that could not be explained by the drag-based model. More specifically, we excluded from the initial dataset of $160$ events $37$ events for which the speed condition was not satisfied, i.e., the mean CME speed does not lie in the range between the initial CME speed and the average solar wind speed and viceversa; i.e., we considered just events such that
\begin{equation}\label{eq:condition}
    v_0\leq \bar{v}\leq w \quad \text{ or } \quad w\leq \bar{v}\leq v_0,
\end{equation}
where $\bar{v} = 1$ AU/TT is the mean CME speed.
Moreover, prior to the network training, the data have been normalized so that the considered quantities would be comparable, and this has been done separately each time on the training, validation, and test set. 

\begin{table}[]
\centering
\caption{CMEs dataset used in this work, where $r_0$ is a fixed parameter and the Travel Time is the final network output and quantity of interest. The last six quantities are the network possible input features.}
\begin{tabular}{lllll}
Name & Notation & Unity & Description & Source \\ \hline \hline 
CME height of eruption  & $r_0$ & km & $r_0 = 20$ $R_{\odot}$, $R_\odot = 6.957\cdot 10^5$ km & - \\
CME time of eruption  & $t_0$ & s & eruption time on the Sun at $r_0$ & \citep{DelMoro2022} \\
CME Time of Arrival & ToA & s & estimated arrival time at $1$ AU & R \& C \\
CME Travel time & TT & s & estimated time between $t_0$ and ToA & \parbox[t]{3.5cm}{R \& C,\\ \citep{DelMoro2022}}\\
\hline
CME initial speed  & $v_0$ & km/s & initial propagation speed from eruption  & LASCO    \\
CME mass & $ m$ &  g & estimated CME mass  &   LASCO   \\
CME impact area  & $A$& km$^2$ & CME impact area, constant angular width &  LASCO\\
Solar wind density & $\rho$ & g/km$^3$ & mean over one hour after $t_0$ & CELIAS \\
Solar wind speed   & $w$ & km/s & mean over one hour after $t_0$ & CELIAS \\
Drag parameter & $C$ & dimensionless & parameter of the drag based model & this work 
\end{tabular} 
\label{tab:dataset}
\end{table}

\subsection{Estimate of the drag parameter}
An accurate estimate of the drag parameter $C$ is crucial for an effective prediction of the TT. Such estimate can be performed according to the following three approaches:
\begin{enumerate}
    \item Following \citep{DelMoro2022}, an analytic inversion of equation (\ref{solution-drag}) for each observed event can be computed.
    \item One can exploit the first neural network in the cascade described in the previous section, by using a subset of the whole data archive as training set.
    \item As in the previous item, one can apply $N1$ but, this time, the training phase is performed using the whole data set at disposal. 
\end{enumerate}

In order to validate the accuracy of the outcome of these three approaches, for each estimated $C$ value we computed $r(C)$ a posteriori, as in equation (\ref{solution-drag}) and compared the result with respect to $1$ AU (specifically, we required $0.95 < r(C) < 1.05$ AU). We found that, using approach 1, i.e., taking the values from \citet{DelMoro2022}, the success rate for this condition is $19.78 \%$; using approach 2, it is between $70-80 \%$ (by varying the subset of the whole data archive as training set); and, finally, using approach 3 it is above $90 \%$. This result is not surprising, since the training process explicitly minimizes the discrepancy between $r(t,C)$ and $1$ AU. Understanding the reason why the network is not able to generalize on the test set still represents an open issue. This problem, which should be addressed in a separate study, is most likely related to the size of the archive at disposal and to the correct balance of the training set \citep{guastavino2022implementation}.


\subsection{TT prediction}
\begin{figure}[h!]
    \centering
     \subfigure[{C1}]{\includegraphics[width=0.45\textwidth]{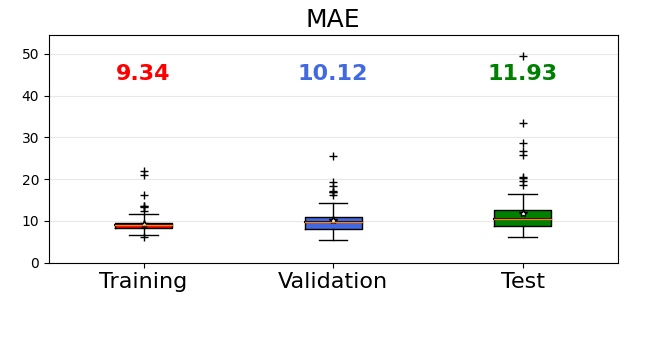}}
         \subfigure[{C4}]{\includegraphics[width=0.45\textwidth]{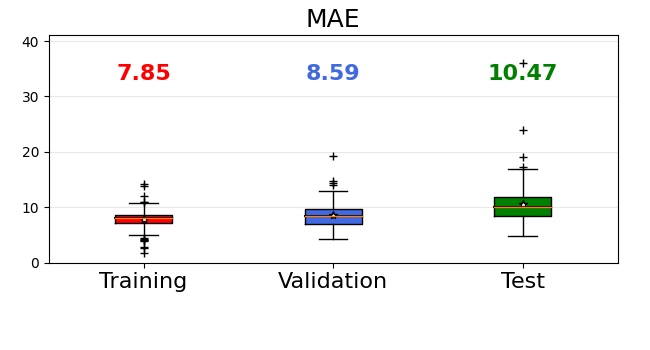}}\\
         \subfigure[{C2}]{\includegraphics[width=0.45\textwidth]{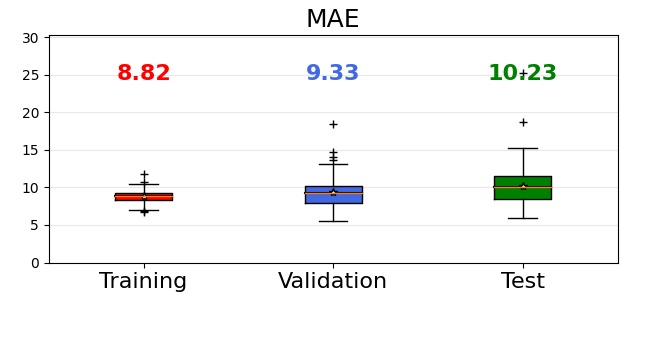}}
     \subfigure[{C5}]{\includegraphics[width=0.45\textwidth]{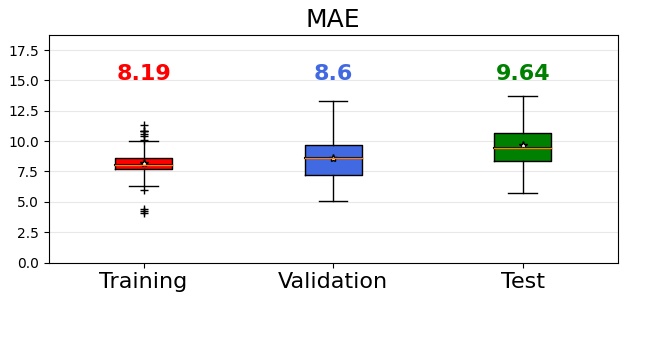}}
      \subfigure[{C3}]{\includegraphics[width=0.45\textwidth]{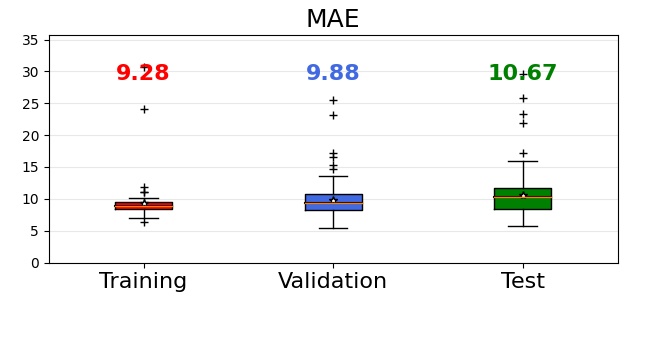}}
     \subfigure[{C6}]{\includegraphics[width=0.45\textwidth]{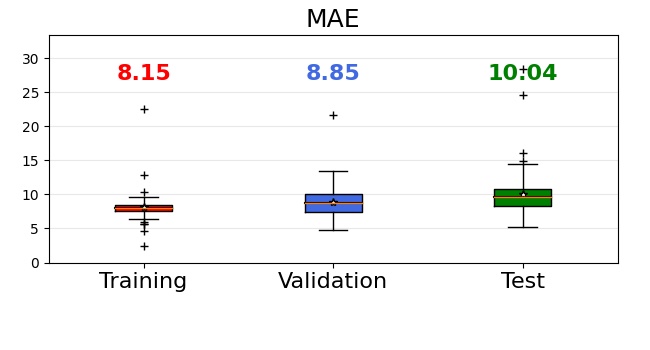}} 
     \caption{Distributions of mean absolute errors for the prediction of TT on 100 realizations of training (red box-plot), validation (blue box-plot) and the test (green box-plot) sets. The numbers in figures represent the mean value of the distributions of MAEs. First row: results with configurations C1 and C4.
     Second row: results with configurations C2 and C5.
     Third row: results with configurations C2 and C5. }
    \label{fig:box-plots}
\end{figure}

\begin{figure}
\begin{center}
\includegraphics[width=7.cm]{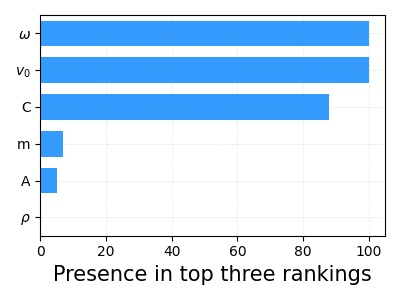}
\caption{Impact of features on network predictions according to permutation importance. The most important features are the wind speed, the CME speed, and the drag parameter $C$.}\label{fig:importance}
\end{center}

\end{figure}

\begin{figure}[h!]
    \centering
     \subfigure[{C1}]{\includegraphics[width=0.45\textwidth]{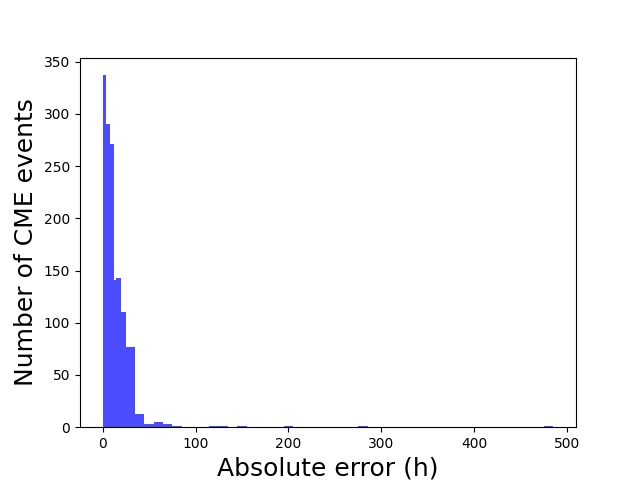}}
         \subfigure[{C4}]{\includegraphics[width=0.45\textwidth]{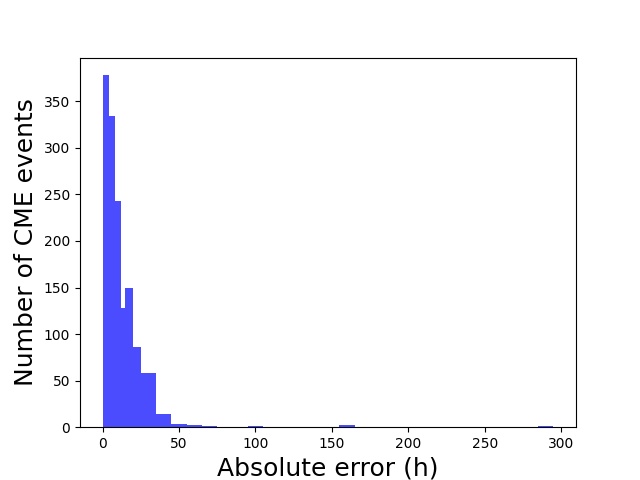}}\\
         \subfigure[{C2}]{\includegraphics[width=0.45\textwidth]{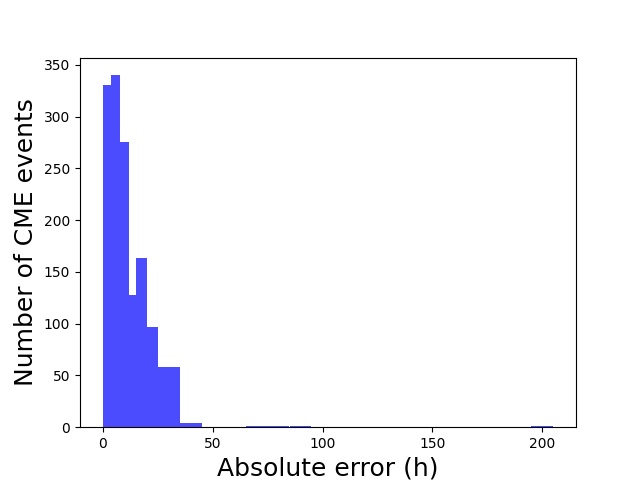}}
     \subfigure[{C5}]{\includegraphics[width=0.45\textwidth]{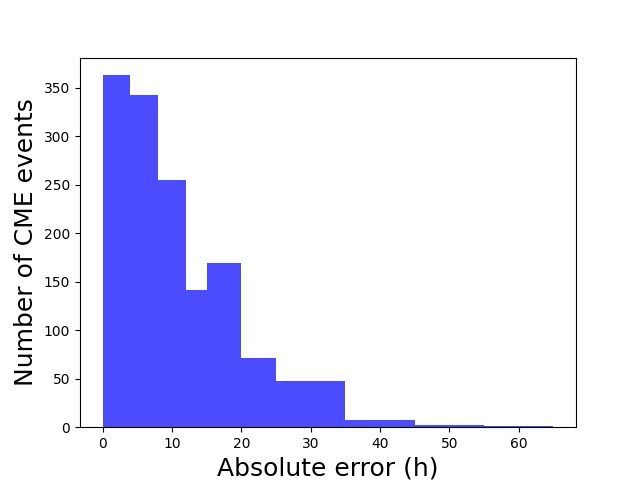}}
      \subfigure[{C3}]{\includegraphics[width=0.45\textwidth]{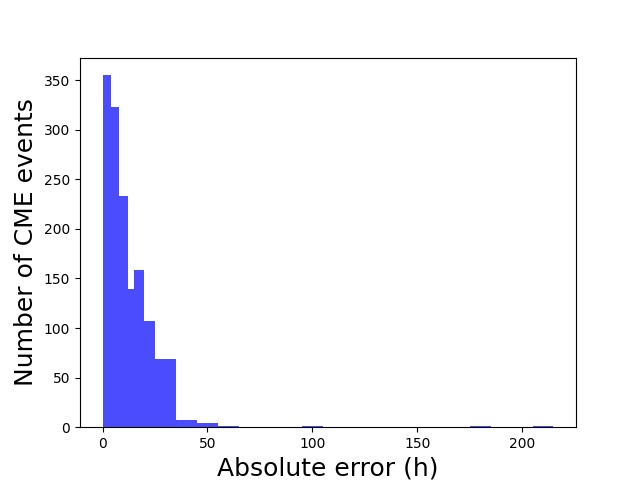}}
     \subfigure[{C6}]{\includegraphics[width=0.45\textwidth]{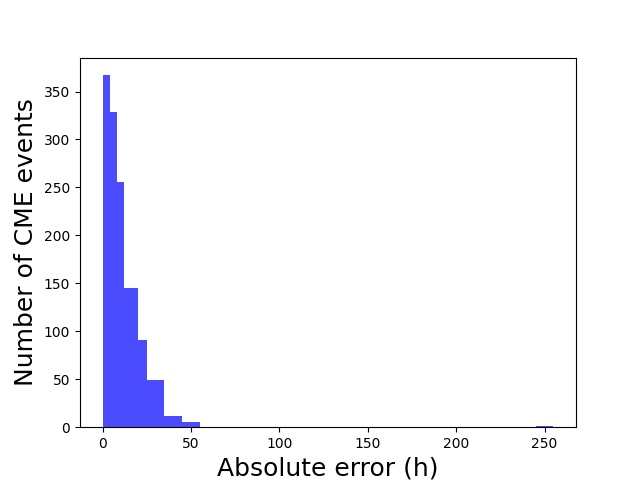}}
    
     \caption{Distribution of absolute errors for the prediction of ToA for all samples in test sets for each configuration. First row: results with configurations C1 and C4.
     Second row: results with configurations C2 and C5.
     Third row: results with configurations C2 and C5. }
    \label{fig:histograms}
\end{figure}

\begin{figure}[h!]
    \centering
     \subfigure[{C1}]{\includegraphics[width=0.38\textwidth]{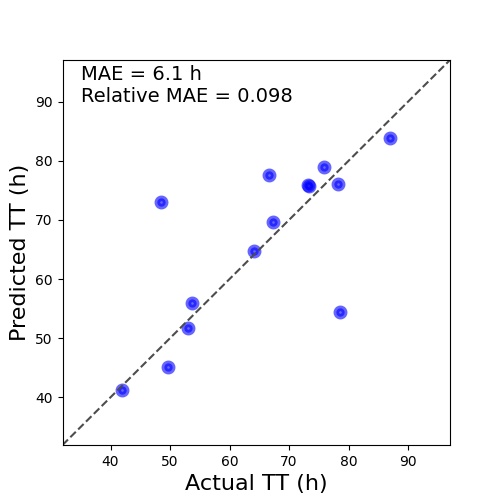}}
         \subfigure[{C4}]{\includegraphics[width=0.38\textwidth]{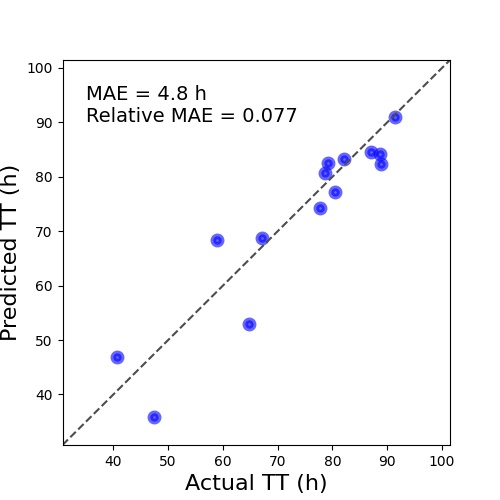}}\\
         \subfigure[{C2}]{\includegraphics[width=0.38\textwidth]{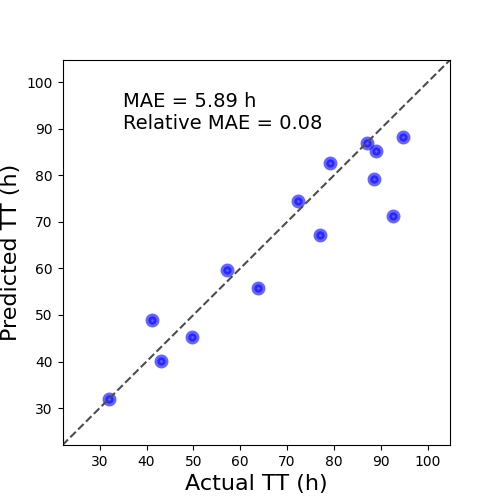}}
     \subfigure[{C5}]{\includegraphics[width=0.38\textwidth]{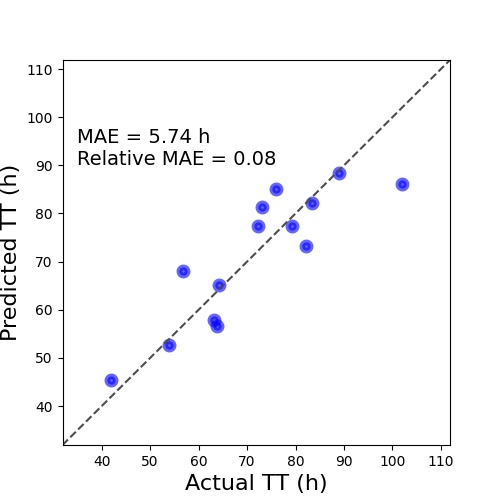}}\\
      \subfigure[{C3}]{\includegraphics[width=0.38\textwidth]{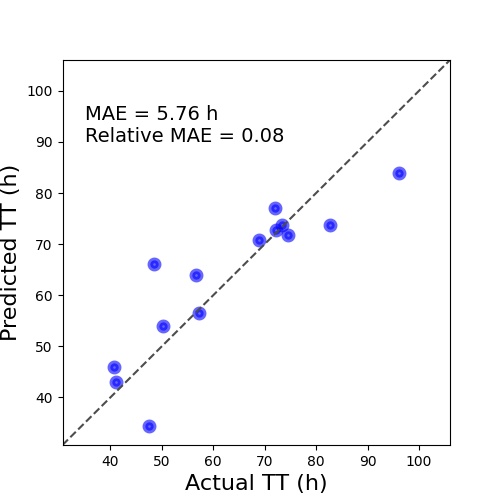}}
     \subfigure[{C6}]{\includegraphics[width=0.38\textwidth]{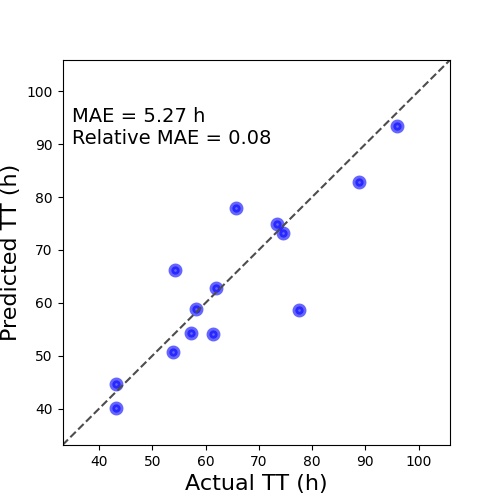}}
    
     \caption{Predicted transit time vs actual transit time for
CMEs in the test set in which the method reached the best performance. First row: results for configurations C1 and C4. Second row: results for configurations C2 and C5. Third row: results for configurations C3 and C6.}
    \label{fig:scatter}
\end{figure}

In order to perform a statistical assessment of this physics-driven AI approach to ToA prediction, we realized $100$ random realizations of the training, validation, and test sets using the $123$ events in our archive made of LASCO and CELIAS observations. Once the networks cascade has been trained, we assessed the prediction performances of the six cascade configurations for each element in the training, validation, and test sets, by comparing the prediction outcomes with the experimental TT values. As a result, Figure \ref{fig:box-plots} contains the box-plots corresponding to the $100$ computed absolute errors and, for each box-plot, the corresponding mean absolute error (MAE). Furthermore, for each realization of the training set we have computed the impact on predictions of each feature according to permutation importance. Permutation importance is computed after a model has been fitted, and it is commonly used to assess how the accuracy of predictions is affected if a single column of the validation data is randomly shuffled \cite{breiman2001random}. Model accuracy is more negatively affected if one shuffles a column that the model heavily relied on for predictions. The results are described in the bar-plots shown in Figure \ref{fig:importance}: for each realization we made a ranking of feature importance and we computed how many times each feature is present in the top three ranking. Then, in Table \ref{tab:results} we reported the minimun (min), mean, median and maximum (max) values of the MAEs on the 100 realizations of the test sets. Figure \ref{fig:histograms} contains the distributions of the absolute errors between the predicted TT and the experimental TT values on all the test samples. The first three bins are 4-hr wide, the fourth bin is between 12 h and 15 h, the fifth and sixth bins are 15 h - 20 h and 20 h - 25 h and the others bins are 10-hr wide.
Figure \ref{fig:scatter} contains scatter plots of the actual and predicted TT values corresponding to the realization of the test set characterized by the minimum MAE. Blue dots represent the different CME events in the selected test set and the
black dashed line represents a perfect prediction, i.e. when the
predicted TT matches the actual TT values. 

\begin{table*}[ht]
\centering
\caption{Minimun (min), mean, median and maximum (max) values of the MAEs on the 100 realizations of the test sets and corresponding relative absolute errors with respect to the observed TT. The best results are in bold face.}\label{tab:results}
{
\begin{tabular}{c | l l l l | l l l l }
\multirow{2}{*}{Configuration} & \multicolumn{4}{c|}{MAE (h)} &  \multicolumn{4}{c}{Relative absolute error } \\
 \cline{2-9}
& min & median & mean & max & min & median & mean & max \\
\hline
C1 & $6.1$ & $10.43$ & $11.93$ & $49.5$ & $0.1$ & $0.16$ & $0.2$ & $1.32$   \\ 
C2 & $5.89$ & $10.03$ & $10.23$ & $25.29$ & $0.08$ & $0.15$ & $0.16$ & $0.6$   \\ 
C3 & $5.76$ & $10.28$ & $10.67$ & $29.63$ & $0.08$ & $0.15$ & $0.17$ & $0.66$  \\ 
C4 & $\mathbf{4.8}$ & $9.96$ & $10.48$ & $36.09$ & $\mathbf{0.07}$ & $0.15$ & $0.17$ & $0.9$   \\ 
C5 & $5.74$ & $\mathbf{9.46}$ & $\mathbf{9.64}$ & $\mathbf{13.75}$ & $0.08$ & $\mathbf{0.14}$ & $\mathbf{0.15}$ & $\mathbf{0.24}$   \\ 
C6 & $5.27$ & $9.59$ & $10.04$ & $28.45$ & $0.08$ & $0.14$ & $0.16$ & $0.72$   \\ 
\hline
\end{tabular}
}
\end{table*}


\section{Comments and conclusions}
The objective of this study was to understand whether and to what extent the use of deterministic information allows machine learning to improve its effectiveness in the forecasting of the CME TT given a very limited number of experimental features at disposal. To this aim, we have exploited the drag-based model as the source of such deterministic information, essentially for the reason why its integrated form can be easily encoded as the loss function applied in the training phase of the AI data-driven approach. The need to optimize the drag parameter in the model inspired a complex architecture made of two neural networks, the first one used to estimate the parameter, and the second one used to predict the CME TT. The possibility to use (or not to use) the drag parameter as input feature for this second network allowed the definition of six configurations for the cascade, and the results in the previous section now allow a comparison of their reliability and robustness.

We first point out that C1 is the only fully data-driven configuration, and it is the one characterized by the worse predicted TT values in the training, validation, and test sets. Further, Table \ref{tab:results} and Figure \ref{fig:histograms} show that this configuration is characterized by several significant outliers. The drag-based model plays an active role in all other configurations, and this increases both the prediction accuracy and its robustness. In particular, the best results are obtained when the loss function has both data- and physics-driven components and $C$ is considered as an input feature for the second network. In general, for each configuration adding $C$ as an input feature improves the predictions. Figure \ref{fig:scatter} shows this behavior in a very clear-cut fashion: the scatter plots in the second column of the figure, which correspond to configurations C4, C5, and C6, respectively, present MAE values that are among the smallest ones one can find in the scientific literature as far as our knowledge is concerned; the same holds for the absolute errors corresponding to the single predictions. However, when the physics-component is present in the loss function good results are obtained also when C is not used as additional input feature: from an operational viewpoint when a new input arrives we can use the second trained network without the need to estimate $C$ for that event.

As said, in configurations C4, C5, and C6, the drag parameter $C$ is used as input feature. The feature importance analysis in Figure \ref{fig:importance} shows that, in these configurations, the drag parameter is among the features that mostly impact the prediction, coherently with the fact that an accurate estimate of $C$ leads to accurate predictions. When $C$ is estimated by $N1$ using the whole archive as training set, the corresponding values of $\gamma=\frac{CA\rho}{m}$ are almost all in the range $10^{-8} km^{-1} < \gamma < 10^{-6} km^{-1}$ observed in \cite{DelMoro2022}. 

Our work-in-progress concerning CME prediction and characterization is currently along three directions. First, we want to generalize this approach to a multi-target version able to provide physics-supported forecasting of both the ToA and the SoA. Second, following the approach proposed in \cite{guastavino2022implementation,guastavino2023operational} we aim at generating balanced training, validation, and test sets in order to account for the data types present in the mission archives.  Third, at a more technical level, we want to update the training phase of the cascade by using probabilistic loss functions designed to optimize specific skill scores \citep{marchetti2022score,guastavino2022bad}.
Moreover, in the future developments of this work we will investigate possible modifications of the drag-based model applied here to include other possible physical phenomena occurring during the ICME propagation, such as the effect of plasma pile-up due to the interaction of the CME flux rope with the surrounding solar wind plasma (an effect that can be taken into account by introducing the so-called ``virtual mass", see e.g. \citet{Cargill2004}, and observed e.g., by \citet{telloni2021study}), and the occurrence of magnetic reconnections with the background interplanetary magnetic field (an effect leading to the so-called ``magnetic erosion" process, see e.g. \citet{wang2018understanding} and \citet{telloni2020magnetohydrodynamic}).

\section*{Acknowledgement}
\label{sec:acknowledge}
SG and FM were supported by the Programma Operativo Nazionale (PON) “Ricerca e Innovazione” 2014–2020; VC by the Universit\`a di Genova within the {\it Bando per l’incentivazione alla progettazione europea (BIPE)} - Mission 1 “Promoting Competitiveness” 2020. VC was also partially supported by INdAM-GNCS. AMM, RS and DT acknowledge the support of the Fondazione Compagnia di San Paolo within the framework of the Artificial Intelligence Call for Proposals, AIxtreme project (ID Rol: 71708). AMM is also grateful to the HORIZON Europe ARCAFF Project, Grant No. 101082164.



\bibliography{refs}{}
\bibliographystyle{aasjournal}

\end{document}